# Dust Collisions in Protoplanetary Disks: Atomic Simulations of the Surface Free Energy

Authors: L. S. Morrissey[1,2,3], D. S. Ebel[3], L. E. J. Eriksson[4,5], A. Georgiou[2], Z. Huang[6], M.-M. Mac Low[4], T. Pfeil[7]

Abstract

Coagulation of dust particles in protoplanetary disks is the first step on the journey to the formation of planets. The surface free energy (SFE) of the dust particles determines the effectiveness of particles sticking to each other after collision, as well as the critical collision velocity above which fragmentation will occur. Studies of SFE have focused on the simplest silicate, silica, usually at standard temperature and pressure. However, protoplanetary dust grains have a wide variety of mineralogical compositions, temperatures, and a low-pressure environment lacking in water vapor. We perform molecular dynamics simulations using a ReaxFF-type potential of the

[1] Corresponding author: Liam Morrissey (lsm088@mun.ca)

[2] Engineering and Applied Science, Memorial University, St. John's, NL, A1B2W4, Canada

[3] Department of Earth & Planetary Science, American Museum of Natural History, New York, NY, 10024, USA

[4] Department of Astrophysics, American Museum of Natural History, New York, NY, 10024, USA

[5] Institute for Advanced Computational Sciences, Stony Brook University, Stony Brook, NY, 11794, USA

[6] Daniel Guggenheim School of Aerospace Engineering, Georgia Institute of Technology, Atlanta, GA, USA.

[7] Center for Computational Astrophysics, Flatiron Institute, New York, NY, 10010

SFE of silica, albite, and anorthite at temperatures ranging from 30 to 700 K in a true vacuum. We find that the SFE drops by tens of percent with increasing temperature or shifting to more complex silicate compositions. More dramatically, we find that the values of the SFE in a vacuum are two orders of magnitude higher than those usually measured in terrestrial laboratories. Our results confirm previous work that suggests that hydroxylation by monolayers of water produces this reduction in SFE in experiments. The coagulation of dust grains thus appears to depend critically on the cleanliness of their surfaces, as well as their temperature and composition.

## 1. INTRODUCTION

The first stage of the formation of planets within a protoplanetary disk is the collision and aggregation of nano- and micro-scale dust particles (Birnstiel 2024; Kimura et al. 2015). This granular aggregation process is typically modelled using Johnson-Kendall-Roberts (JKR; Johnson et al. 1971; Dominik & Tielens 1997) contact theory, a well-established theoretical approach to approximate the sticking, bouncing, and aggregation of colliding grains. (Note that recent simulation work by Yoshida et al. (2024) suggests that large grains are less elastic than suggested by JKR theory.) The behaviors predicted by JKR models are largely determined by the surface free energy (SFE) of the colliding grains (Chokshi et al. 1993; Dominik & Tielens 1997; Kimura et al. 2015; Wada et al. 2009), a model input that is specific to the composition of the grains and local environment during collision. The SFE is a measure of the energy difference between a free surface and the bulk of material, governing the properties of colliding dust grains at their surface interfaces. Grain surfaces with high SFEs adhere more strongly to each other, promoting particle aggregation and thus allowing the growth of larger bodies. In contrast, grain surfaces with low SFE adhere less strongly, making it more difficult for particles to aggregate and grow.

However, the SFE of colliding grains is not well understood for the specific conditions found in protoplanetary disks. Experimental studies on SFEs for pure silica ($SiO_2$) have measured SFE values ranging from 0.02–2.5 J m$^{-2}$, over two orders of magnitude in difference (Blum & Wurm 2000; Kimura et al. 2015). These experimental silica results are typically limited to specific temperatures, often ~300 K. The low values come from experimental surfaces that are likely hydroxylated by ambient water found in the terrestrial environment, while the higher values come from experiments designed to avoid hydroxylation. Furthermore, colliding dust particles in the protoplanetary disk typically contain more elements than just Si and O. They are instead formed, depending on the local disk temperature, from refractory metals and oxides, more complex silicates, organic materials, or ices (Ebel et al. 2006; Grossman 1972; Lewis 1972)). It is unknown how the SFE changes for the different mineral compositions found within the protoplanetary disk. This large uncertainty in SFEs leads us to study SFEs relevant to grain collisions in the protoplanetary disk.

As an alternative to experimental studies, molecular dynamics (MD) modelling of these surfaces on the atomic scale allows energy differences between the bulk and the surface to be calculated as a function of temperature and composition. MD modelling uses an interatomic potential to dynamically compute many-body interactions between all atoms in the system. While this method has been used previously for silica SFEs, inconsistencies exist in the modeling methods used to prepare and relax the exposed surface (Erhard et al. 2024; Nietiadi et al. 2020; Rimsza et al. 2017a). Moreover, results are needed for the temperatures, environments, and mineral types important in protoplanetary disks. Mineral-specific values are essential for models of planet formation and disk evolution, while also influencing our understanding of the chemical processes occurring in the early Solar System.

In this study, we use MD simulations to study the SFEs of crystalline silica, albite, and anorthite; fundamental silicates thought to be relevant for many planetary bodies. We study SFEs for clean surfaces across the complete range of temperatures expected within the protoplanetary disk. These mineral and temperature specific SFEs can be incorporated into existing JKR-based models to improve our understanding of grain accumulation within the protoplanetary disk. We discuss the implications of our newly derived SFEs for key collision behaviors using JKR theory.

## 2. METHODOLOGY

### 2.1 Molecular Dynamics Simulations

MD simulations were performed using the Large-scale Atomic/Molecular Massively Parallel Simulation (LAMMPS) package (Plimpton 1995; Thompson et al. 2022). The accuracy of MD simulations strongly depends on the validity of the interatomic potential used to calculate forces between neighbouring atoms prior to iterating their positions and velocities. Here, we use a ReaxFF-type potential (Van Duin et al. 2001) to model all interactions, allowing for dynamic bond formation and breaking in a multielement substrate and, for future studies, chemical reactions. ReaxFF is uniquely capable of simulating bonded and nonbonded interactions, allowing for both long- and short-range bond contributions to be calculated at each time step. Connectivity-dependent reactions such as valence and torsion energy are modeled so that when bonds are broken their contribution to the total energy is zero. Nonbonded, van der Waals, and long-range Coulomb interactions (cut-off at a standard distance of 10 Å) are calculated irrespective of the connections between all atom pairs in the simulation. For our study, we selected a potential initially developed by Pitman & Van Duin (2012) for silicates and zeolites. This potential has also been extensively validated, showing its ability to predict key properties of silicates with different atomic arrangements and compositions containing C, H, O, Fe, Cl, Si, Al, Ca, and Na, beyond those for which it was initially parameterized for (Yu et al. 2017). Based on these findings, this potential has been used extensively to study key bulk and surface properties for albite and anorthite (Mayanovic et al. 2023; Morrissey et al. 2022, 2024), making it a useful choice for these simulations.

We have conducted MD simulations to study the SFEs of three key minerals relevant to grain collisions within protoplanetary disks. Silica was chosen as it is a commonly studied mineral in previous theoretical and experimental studies on SFEs. Albite ($NaAlSi_3O_8$) and anorthite ($CaAl_2Si_2O_8$) were chosen as they are the two endmembers in the plagioclase feldspar family, a class of silicate minerals thought to be abundant on many celestial bodies (Domingue et al. 2014; McClintock et al. 2018; McCoy et al. 2018; Papike et al. 1991).

## 2.2 Preparing the Bulk Mineral Structures

For each mineral we first developed crystalline bulk samples by replicating the conventional unit cell of each mineral type from the Materials Project database (Jain et al. 2013) to develop a sample containing ~2500-5000 atoms. Boundary conditions were periodic along the x, y, and z directions, simulating an infinite bulk. This sample was then equilibrated to the prescribed temperature using a Berendsen barostat algorithm (Berendsen et al. 1984) which allows the volume of the bulk to change. Temperatures of 30, 100, 300, and 700 K were simulated, capturing the midplane temperature profile of protoplanetary disks (Dullemond et al. 2006) as well as overlapping with a typical experimental temperature. After equilibration for 25 ps, each sample reached a stable temperature and pressure. Each substrate then was brought into charge equilibration using the Electron Equilibration Method approach (Mortier et al. 1986) as implemented in LAMMPS and described by Van Duin et al. (2001). The equilibration minimizes the electrostatic energy by adjusting partial charges on individual atoms based on neighbor interactions. Finally, the potential energy $PE_{bulk}$ for each bulk periodic equilibrated mineral was stored.

### 2.3 Calculating Surface Free Energies

For each of the bulk samples studied, the simulation domain was extended 100 Å in the $z$ direction, creating a free surface exposed to vacuum. Each slab was oriented such that the perfect (001) cleavage plane was exposed perpendicular to the $z$ direction. Boundary conditions in the $x$ and $y$ directions were left periodic whereas in the $z$ direction they were changed to 'fixed', creating an infinite slab. Each of the mineral slabs was then equilibrated in a Berendsen thermostat for 200 ps at the desired temperature, ensuring the surface was adequately relaxed. This relaxation period was longer than several previous MD studies of SFE. The SFE

$$\gamma = \frac{PE_{bulk} - PE_{slab}}{A} \quad (1)$$

was calculated every 5 ps by taking the difference between the stored potential energy of the periodic bulk sample $PE_{bulk}$ and its corresponding slab $PE_{slab}$, divided by the area of the surface $A$.

## 3. RESULTS

### 3.1 Surface Free Energies

Figure 1 displays $\gamma$ as a function of slab relaxation time for each of the different minerals and simulated temperatures. In all cases the SFE converges to a steady value; however, the relative decrease in the SFE with relaxation time depends on temperature. At low temperatures there are only minor differences of < 5% between the initial unrelaxed and final relaxed SFE after 200 ps, whereas at higher temperatures (300 – 700 K) these differences increase to 10-15%.

When a free surface is opened to vacuum several undercoordinated dangling bonds are created. These dangling bonds are energetically unfavorable and increase the SFE of the slab. When the temperature during relaxation is increased these surface atoms have more kinetic energy, allowing them to reorder with time into a more stable and lower energy state. Therefore, it is important that MD studies of SFE allow for sufficient relaxation time as shown in Figure 1.

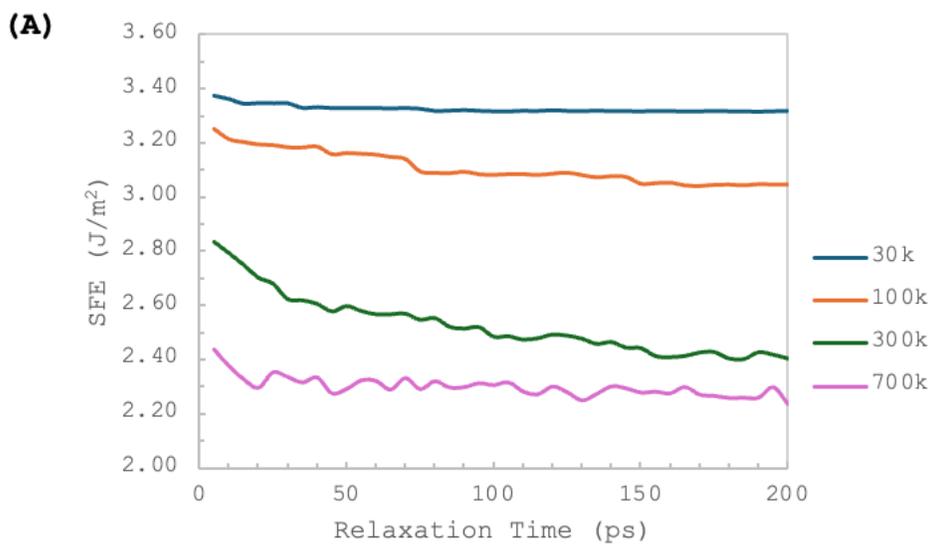

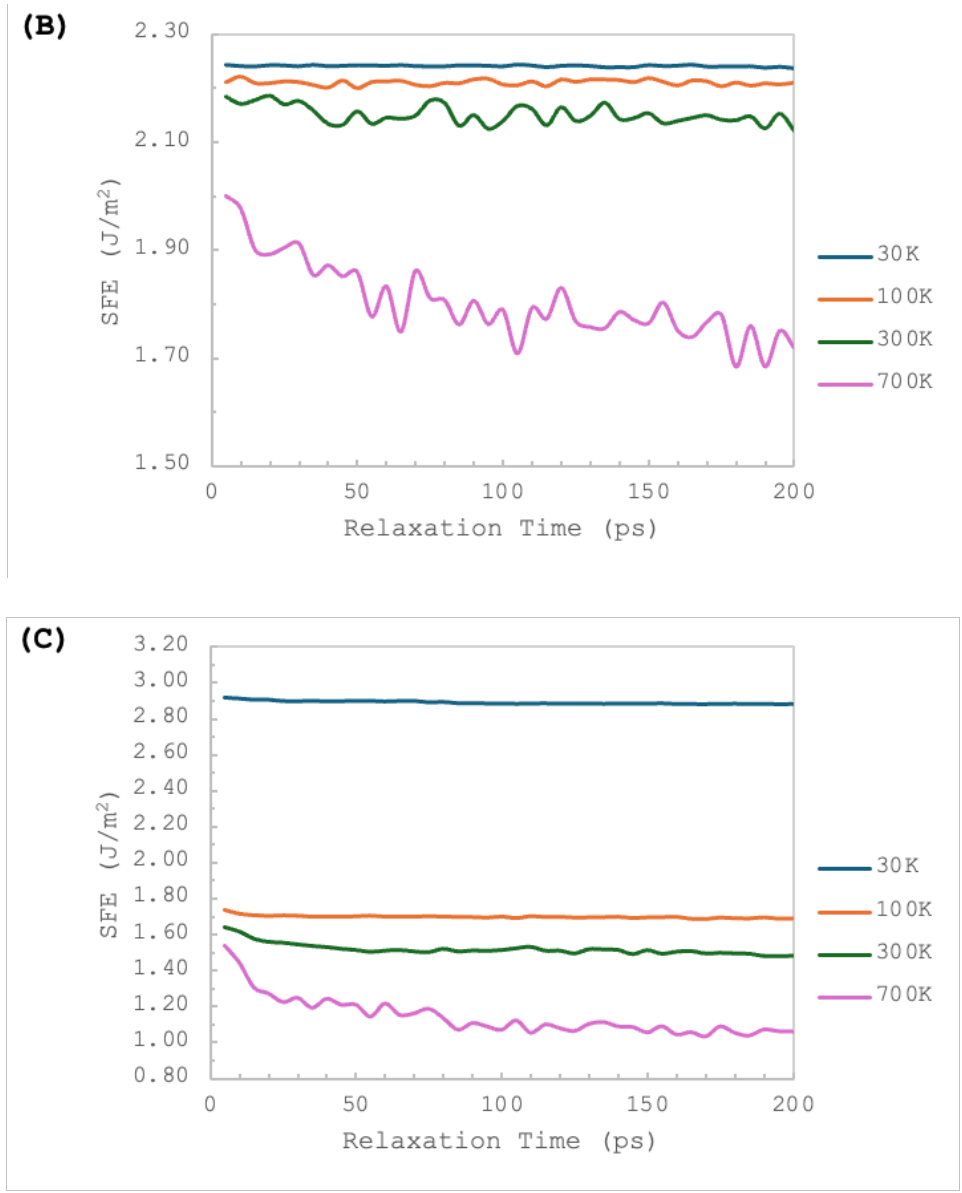

Figure 1: Surface free energies for silica (A), albite (B) and anorthite (C) as a function of relaxation time and temperature

Table 1 shows the final SFE for each mineral type as a function of simulation temperature. There is a clear effect of temperature on the final relaxed SFE for each mineral. In all cases the SFE decreases with temperature. For example, moving from 30 K to 700 K, the SFE decreases by factors of 1.5, 1.3, and 2.7 for silica, albite, and anorthite, respectively. The largest decrease with temperature occurred for silica, which also had the highest SFE and thus most energetically unfavorable surface. Therefore, when simulation temperature was increased this surface rearranged into a more favorable state. At low temperatures there was not enough energy for rearrangement. Figure 2 and 3 shows an example of a relaxed albite and anorthite surface at 30 K and 700 K. For anorthite at 30 K the surface is crystalline and repeating, exhibiting several non-bridging dangling oxygen atoms (shown in red) bound only to aluminum (shown in grey), leading to an increased SFE. At 700 K there is sufficient energy for these dangling oxygen atoms to rearrange into a bridging state and thus reduce the SFE. Similarly, for albite the cleaved surface has several high energy dangling oxygen which can rearrange at 700K into less energetic positions thus reducing the SFE. Future JKR models that use SFE to simulate grain collisions should therefore consider and implement a temperature dependent SFE.

Finally, the SFE is also dependent on the composition of the mineral at each temperature. For all temperatures silica has the highest SFE, followed by either albite or anorthite, suggesting that approximating grains as a simple silica structure may overestimate the SFE. In general, the lowest SFEs are found for anorthite. For example, at 700 K the SFE of silica (2.24 $J/m^2$) is a factor of 2.1 higher than anorthite (1.06 $J/m^2$). We suggest that the source of these differences is in the unique bond types formed by the plagioclase feldspars. Oxygen on silica surfaces is bound only to silicon, forming strong covalent bonds that create a high energy surface. In contrast, in feldspars, Al, Na, and Ca all substitute for the Si, forming comparatively weaker bonds with surface oxygen (Dana 2023). These weaker bonds reduce the energy needed to form a surface, thus lowering the SFE for albite and anorthite.

Table 1: SFE for each mineral type as a function of temperature

| Temperature (K) | Surface Free Energy (J/m$^2$) | | |
| --- | --- | --- | --- |
| | Silica | Albite | Anorthite |
| 30  | 3.32 | 2.24 | 2.88 |
| 100 | 3.04 | 2.21 | 1.70 |
| 300 | 2.40 | 2.11 | 1.47 |
| 700 | 2.24 | 1.71 | 1.06 |

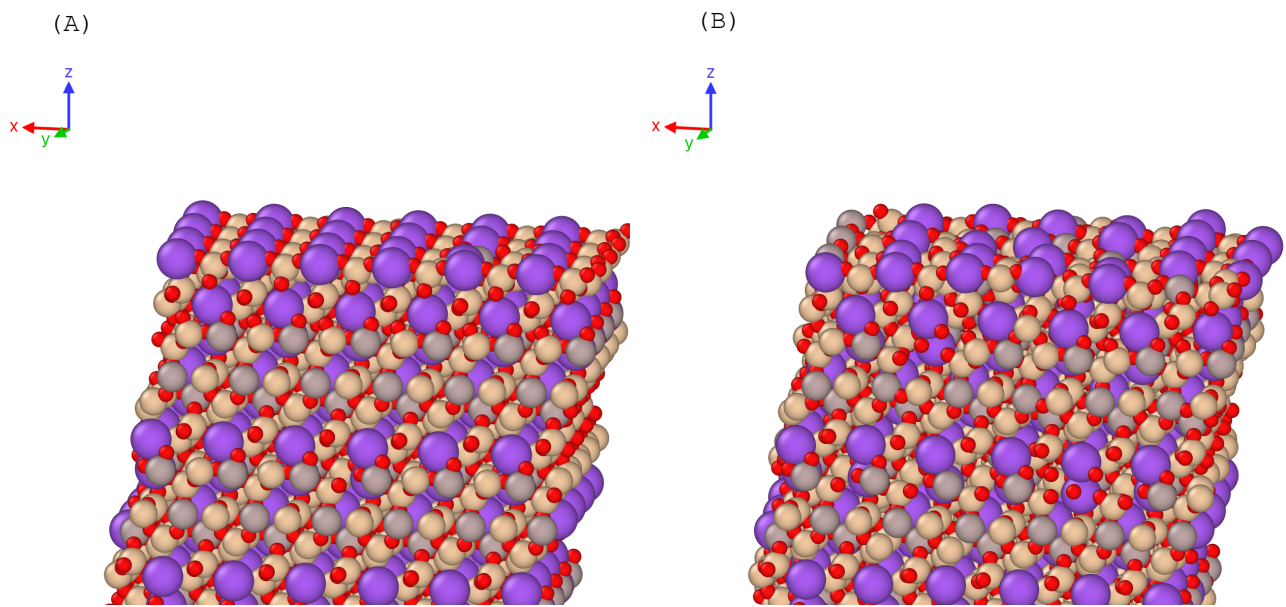

Figure 2: Albite slab at 30 K (A) and 700 K (B). In order of increasing size, oxygen atoms are in red, silicon atoms are in tan, aluminum atoms are in grey, and sodium atoms are in purple.

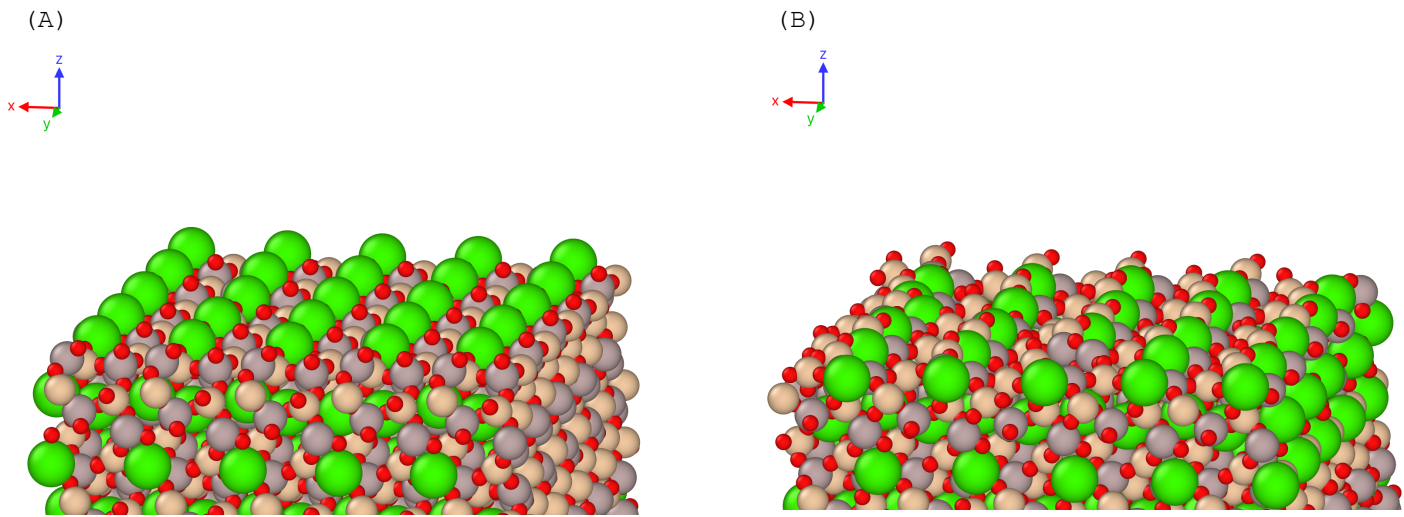

Figure 3: Anorthite slab at 30 K (A) and 700 K (B). In order of increasing size, oxygen atoms are in red, silicon atoms are in tan, aluminum atoms are in grey, and calcium atoms are in green.

3.2 Comparison to Previously Reported Values

3.2.1 Theoretical Modelling

While MD studies on SFEs of albite and anorthite are lacking, several previous simulations have used MD and density functional theory to study SFEs of crystalline and amorphous silica. These studies have reported crystalline silica SFEs of 1.96-2.4 J m$^{-2}$ for relaxed surfaces (Bandura et al. 2011; Grillo et al. 2012; Murashov & Demchuk 2005; Rimsza et al. 2017b), increasing to 4.5-8.4 J ms$^{-2}$ for unrelaxed surfaces (Goniakowski & Noguera 1994). Specifically, Rimza et al. (2017) demonstrated the importance of annealing temperature, showing decreased SFEs with increased annealing temperature. However, these results were limited to amorphous substrates and all samples were cooled to ~0 K after annealing and before SFE calculation. More recently, Erhard (2024) used a machine learning potential in MD to study SFEs of (001) crystalline silica, predicting SFEs of ~4.0 and ~3.0 J m$^{-2}$ for unrelaxed and relaxed substrates, respectively. Overall, our results at these temperatures agree well with the body of previous simulations on the SFE of silica and the importance of relaxation, validating our methodology and our choice of interatomic potential.

### 3.2.2 Experimental Studies

In contrast to the theoretical values, there is a large body of literature giving experimental values for the SFE of crystalline and amorphous silica, which find typical values from 0.02–0.025 J m$^{-2}$ (Blum & Wurm 2000; Kimura et al. 2015). Thus, there is a two order of magnitude discrepancy between experiment and theory. To explain this, Kimura et al. (2015) highlighted the role of surface hydroxylation during sampling. When silica surfaces are exposed to ambient terrestrial conditions they become quickly terminated with Si-OH groups, leading to the production of several monolayers of $H_2O$. This environmental hydroxylation covers the dangling O bonds on the surface, significantly reducing the SFE. While ambient $H_2O$ is ubiquitous on Earth with abundances approaching 0.01 by mass, it is far less common within the protoplanetary disk gas. In regions within the water ice line, where ice has been sublimated, water abundances in the gas may reach as high as $10^{-5}$ (Harsono et al. 2020), while outside the ice line surfaces are likely to be coated with ice. The major constituent of protoplanetary disks is less-reactive, low-density molecular hydrogen with typical densities of $10^{-8}$-$10^{-11}$ g cm$^{-3}$. When surfaces were studied in near vacuum conditions, where coverage of <1 monolayer of $H_2O$ was present, SFEs increased an order of magnitude to 0.2 J m$^{-2}$ (reviewed by Kimura et al. 2015). When experimental surfaces were deliberately heated, thus increasing the desorption of volatiles, or studied in anhydrous environments, SFEs increased by over two orders of magnitude to 1-3 J m$^{-2}$ (Kimura et al. 2015 and references therein). Building on this work Pillich et al. (2021) compared relative SFEs of samples with many monolayers of $H_2O$, one monolayer of $H_2O$), preheated to remove all $H_2O$. They found an order of magnitude increase in SFE from many to single monolayers samples, followed by another order of magnitude increase from a single monolayer to completely dry (which they characterize as 'super-dry'). They conclude that this large increase from the entire removal of $H_2O$ has yet to be included in models of protoplanetary disks and requires more study. It is important to note that these studies were focused on amorphous silica but noted the potential importance of crystalline inclusions in grains, which have received less focus. For the present study, our surfaces in MD were in perfect vacuum with zero OH or $H_2O$ coverage. We support the conclusions of Pillich et al. and build on their work by directly quantifying the SFE of completely dry silicates. Our work highlights the importance of considering the temperature and ambient environment of the colliding grains before selecting an SFE. Future research is needed to study the effects of ambient molecules at different intermediate coverage rates for the minerals and temperatures considered here.

### 3.3 Implications for JKR Modelling in Protoplanetary Disks

To quantify the importance of the SFE in predicting dust collision behavior we calculate the fragmentation velocity $v_{frag}$, the velocity at which a grain/particle breaks apart upon collision, for the various simulated SFEs. Following the work of Wada et al. (2009), and adjusting for the correct dependence of $a_0$ on $r$ (Chokshi et al. 1993), we derived

$$v_{frag} = 18.7\ m\ s^{-1} \left(\frac{\gamma}{0.1\ Jm^{-2}}\right)^{\frac{5}{6}} \left(\frac{r}{0.1\mu m}\right)^{-\frac{5}{6}} \left(\frac{E}{7 GPa}\right)^{-\frac{1}{3}} \left(\frac{\rho}{1000\ kgm^{-3}}\right)^{-\frac{1}{2}}, \quad (2)$$

where Young's modulus is $E$, the grain radius is $r$, and the grain density is $\rho$.

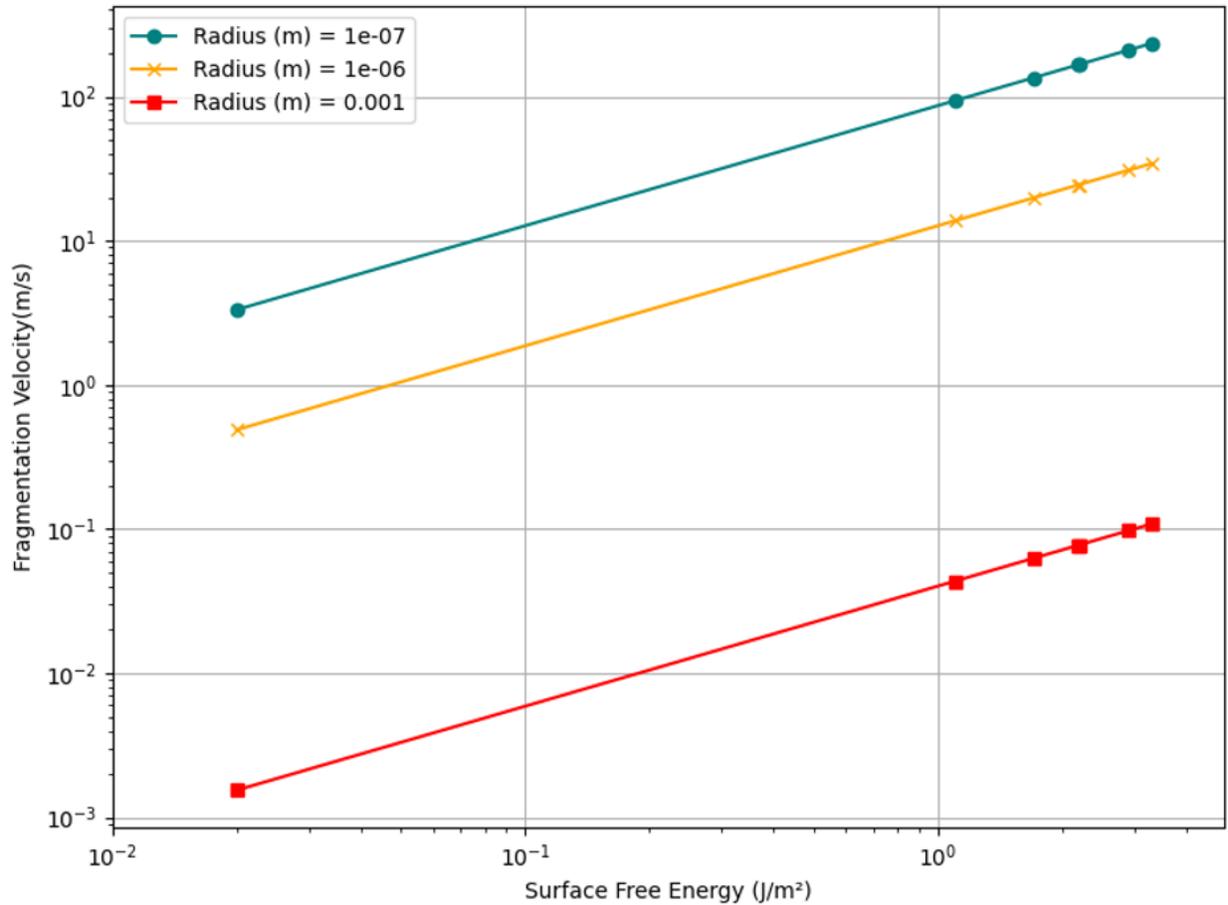

Figure 3: The effect of SFE on the fragmentation velocity. We assume identical silica grains with radii given in the legend colliding with an elastic modulus of 70 GPa, and a grain density of 2200 kg m$^{-3}$.

Here we considered grain radii from 0.1 mm – 1 mm, capturing the range of sizes expected within the protoplanetary disk. In all cases the predicted fragmentation velocity from JKR modelling is highly influenced by the variation in SFEs we have described. First, for the role of composition and temperature, as we move from the lowest (anorthite at 700 K) to the highest calculated SFE (silica at 30 K) the fragmentation velocity increases by a factor of 2.5. Therefore, accounting for grain composition and temperature has an important effect on predicted collision behavior. Further, when the SFE for hydroxylated silica derived from experiment is used (0.02 J m$^{-2}$) the fragmentation velocity is reduced by a factor of 28-70, almost two orders of magnitude different from $v_{\text{frag}}$ for clean samples and significantly higher than the effects of composition and temperature. This further highlights the need to better understand hydroxylation rates on silicate surfaces and their effects on SFEs. In addition to fragmentation velocity, the SFE also influences other important quantities for grain interactions in aggregates such as the energy for starting and continuing rolling and twisting and the sliding friction (Dominik & Tielens 1997).

## 4. CONCLUSIONS

Our novel results highlight the key role MD simulations can play in better understanding grain collisions within protoplanetary disks. We show that SFEs of minerals are temperature and mineral specific, and support findings that SFEs are highly sensitive to $H_2O$ in the environment. Our simulated SFEs for clean, dry surfaces were significantly higher than commonly used values in JKR models of grain interactions in protoplanetary disks. These studies typically use only a single value of the SFE that derives from experiments conducted on silica in ambient terrestrial conditions. Such approximations can introduce important errors into JKR modelling of collisional behavior, strongly affecting our understanding of dust evolution in protoplanetary disks.

For example, moving from the commonly approximated silica to more complex minerals like albite and anorthite decreases the SFE by factors of order unity in all cases, as does increasing the ambient temperature. On the other hand, moving from commonly used experimental values at standard temperature and pressure in Earth atmosphere to clean surfaces possibly more appropriate for the low-pressure molecular hydrogen environment of a protoplanetary disk can increase the SFE by two orders of magnitude, dramatically increasing the velocity at which grains can collide without fragmenting.

Therefore, SFEs that account for the composition of the colliding grains and local environment during collision should be used in coagulation models. Here, we provide an initial set of SFEs for clean crystalline silicates across the range of expected temperatures in the protoplanetary disk. In future work we will build on this study by considering the effects of amorphization, other minerals such as olivine, and surface coverage by different molecules.

## Acknowledgments

We thank S. Verkercke and J. Drążkowska for useful discussions and Xiao Ping Zhang for quick confirmation of the consistency of Chang'E bouncing dust results with SFEs from contaminated surfaces despite a unit problem in the publication. LM and JR were supported in part by the NSERC Discovery Grant and the CSA Research Opportunities in Space Sciences program awards. This research was supported by the International Space Science Institute (ISSI) in Bern through the ISSI International Team project #616 "Multi-scale Understanding of Surface-Exosphere Connections (MUSEC). LE and M-MML acknowledge support from NASA Emerging Worlds grant 80NSSC25K7117. This research has made use of NASA's Astrophysics Data System Bibliographic Services.